# Learned Cost Model for Placement on Reconfigurable Dataflow Hardware

Etash Guha[1], Tianxiao Jiang[2], Andrew Deng[3], Jian Zhang[4] and Muthu Annamalai[5]

SambaNova Systems
Palo Alto, CA, USA

**Abstract**: Mapping a dataflow-graph of an ML model onto a reconfigurable system is difficult, as different mappings have different throughputs and consume resource constraints differently. To solve this, a model to evaluate the throughput of mappings is necessary as measuring throughput completely is expensive. Many use a hand-designed analytical model, relying on proxy features or intuition, introducing error. We provide a Learned Approach that predicts throughput 31%-52% more accurately over a variety of graphs. In addition, our approach shows no accuracy degradation after removing performance annotations. We show that using this approach results in 5.6% faster compiled graphs.

Index Terms—Data-driven, cost model, dataflow architecture

## I. INTRODUCTION

For modern deep neural network (DNN) models [4], [12], efficient model training requires both high compute capacity and high memory bandwidth from hardware. To reach high hardware efficiency in model training, reconfigurable dataflow architectures are being increasingly adopted for building next-generation training accelerators [3], [11]. In these dataflow architectures, a large array of reconfigurable function units, including both compute and memory units, are interconnected on-chip as shown in Figure 1a. With an intelligent compiler, such architectures can achieve high compute capacity and memory bandwidth at the same time by leveraging the interconnected on-chip functional units.

To enable DNN training on dataflow architecture, compilers first extract a directed acyclic graph (DAG) from DNN with common arithmetic operations such as matrix multiplication as nodes. These nodes are then placed onto the functional units and routed through on-chip interconnections; this process is known as placement and routing (PnR). To determine the optimal PnR decision that maximizes the throughput of the compiled DNN, compilers optimize a cost model which ranks the fitness of different PnR decisions. For this process, a precise cost model for PnR is critical to the success of compilers for dataflow architectures.

It is a common choice to construct cost models for PnR of dataflow architectures based on heuristics involving placement density and routing congestion [2]. These heuristics model functional unit operation and interconnect transmit speed to estimate hardware throughput during operation. However, establishing functional heuristics that cover each arithmetic operation presented in DNNs can be iterative and time consuming. Heuristics also lack precision in modeling complex empirical behaviors of the functional units and on-chip interconnects; this could lead to over-pessimistic or over-optimistic cost modeling. Finally, heuristic rules are often not adaptive enough to suit continuous development due to substantial adhoc tweaking needed to accommodate changes in the compiler.

Authors 1, 2 have equal contribution. Authors 3, 4 contributed to the validation and methodology. Author 5 conceptualized and bootstrapped the data and effort.

In this paper, we challenge the common belief in the necessity of heuristics in cost modeling for placement and routing in dataflow architectures. We ask, "can we learn a cost model purely from empirical measurements using machine learning techniques, and achieve superior modeling precision, less engineering time cost and stronger adaptivity compared to heuristic-based cost models on real DNN workload ?"

Concretely, we present a data-driven cost model based on graph neural networks (GNN) [14] for the dataflow architecture. In GNN, nodes and edges are represented by real-valued embedding vectors to encode local information. By iteratively aggregating the local information associated with the nodes and edges across the graph neighborhoods, GNN can capture the holistic characteristics of the entire graph structure. We leverage the GNN to extract the PnR graph representation and use a regressor to learn the measured throughput. In this way, the cost model can gain throughput prediction capabilities and also generalize to different unseen graphs by learning across broad set of PnR decisions.

We design the data-driven cost model to resolve three pain points from heuristic-based cost modeling:

- • High engineering time cost on complex heuristic derivation. Our GNN-based cost model does not require complex heuristics, for example, the latency of the functional unit on performing specific arithmetic operations and the detailed communication latency between the units.

- • Poor accuracy in predicting the real performance. Our cost model learns directly from empirical measurements. Additionally, it is able to capture subtleties in hardware behaviors which are hard to encode by rigid rules.

- • Ad-hoc tweaking in heuristics under continuous compiler development. For our cost model, adaptivity in the continuous development can be achieved quickly by recollecting empirical throughputs and retraining the GNN regressor within hours.

Empirically, we first show that our data-driven cost model can more precisely rank the throughput of PnR decisions than heuristic baselines. We then swap in our cost model into the fast-evolving compiler for an industry-leading training accelerator [1]. Using this compiler, we demonstrate that the artifacts compiled with our data-driven cost model consistently deliver higher throughput at different timepoints, showing adaptivity to the continuous changes in the compiler stack.

More specifically, on a randomly generated (training) set of PnR decisions, we show that the throughput predictions from our cost model can achieve up to 20% higher Spearman rank correlation coefficient on the empirical throughput than the heuristic-based baselines across all datasets. To validate the quality of our cost models in compiling large real DNNs workload, we integrate our cost model into a production-level compiler. We demonstrate that the PnR decisions optimized on our data-driven cost model achieve up to 5% higher training throughput than the decisions driven by the default heuristic-based cost model across transformer based DNNs such as Bert-large and GPT2 models [4], [12]. By repeating the data collection and model training at two time points where there is a major compiler update, we show that both BERT and GPT compiled with our data-driven cost model consistently has higher throughput compared to the baselines.

Our contributions are summarized as follows.

We propose a purely data-driven cost model without heuristic rules for the PnR process on dataflow architectures. This cost model can be learned within hours instead of months for the conventional heuristic-based cost model.

- We show that our trained cost model can reduce the prediction error by almost half testing on a number of building blocks of modern DNNs.
- We empirically evaluate our pretrained cost model by using it to guide PnR to compile BERT on a dataflow architecture and demonstrate better throughput compared to the heuristic-based cost model.
- We demonstrate that our cost model is adaptive to the evolution of compilers by yielding higher throughput consistently on BERT and GPT at two different time points when the compiler is upgraded.

## II. PRELIMINARY

To establish the technical background, we first present the preliminaries on PnR in compilers for reconfigurable dataflow architecture in Section II-A. We then discuss limitations of heuristic-based cost modeling in PnR with examples motivating our study on data-driven approaches in Section II-B. In order to expose concepts for the backbone in our data-driven cost model in Section III, we additionally discuss graph neural networks in Section II-C.

### A. Placement and Routing for Dataflow Architectures

As shown in Figure 1a, hardware using the dataflow architecture usually holds a large number of on-chip functional units including compute units and memory units. Data can be efficiently transmitted in between the units via fast interconnect. This architectural design can deliver both high compute capacity and high memory bandwidth. This makes dataflow architecture a natural fit to breakthrough the efficiency of training DNNs [11], [13].

a) Placement and Routing for Pipeline Execution: To compile a DNN for the dataflow architecture, compilers map the arithmetic operations in the DNN to the reconfigurable functional units; this process is usually termed as placement and routing (PnR). In this process as shown in Figure 1b, compilers abstract the DNN into a dataflow graph and place all the arithmetic operations onto the reconfigurable functional units.1 After placing the operations, the compiler determines routes to communicate the data across the units. This spatial mapping keeps data flowing on chip and enables high utilization of the hardware. In more detail, modern DNNs are trained using iterative algorithms where a batch of training data samples get processed at each iteration. Using the spatial mapping, data samples in a batch flow through the operations in pipeline fashion. For example, in Figure 1b, the linear and ReLU can be two different pipeline stages and they can process different samples at the same time. At steady state, the pipeline execution keeps all functional units utilized concurrently.

b) Cost modeling: To determine an optimal PnR decision on the dataflow graph, compilers need to search a large solution space. This is similar to the NP-hard physical cell placement problem in VLSI design [8]. To achieve a practical solution, dataflow architecture compilers use a heuristic based cost model which measures the fitness of PnR decisions to guide the solution space search. This cost model can be used by a placer algorithm, for example simulated annealing [8], to iteratively optimize the PnR solution. The precision of this cost model is critical to compiling DNNs with high training throughput.

### B. Limitations of Heuristic-based Cost Models

Generally, it is hard to derive a rule-based analytical cost model that captures all the subtleties in a PnR decision on a large array of functional units with complex communication patterns. Heuristic rules are used to approximate the computational latency of functional units on each operation, and the data transmit latency between the units. These heuristic rules typically model the local behavior of units for a standalone operation without considering the interactions between them; these heuristics simplify the modeling. However, they can significantly hold back industrial compiler engineering in three important aspects. Firstly, writing the heuristic-rules with acceptable precision to cover the diverse set of operations involved in training DNNs is an incredibly time- consuming effort. Secondly, the imprecision of heuristics can preclude PR decisions which are empirically performant but discouraged under simplified heuristics. For example, two operations may share the shortest path for communication on the fabric, and each could require the full bandwidth. Even though in the dataflow graph context they could time-share the routes at runtime, some conservative heuristics could seek to prevent route congestion and encourage a longer route for one of the operations. Lastly industrial compilers are evolving daily in engineering practice. When substantial changes go into the compiler stack such as upgrades to the low-level implementation of an operation, the heuristic rules require ad-hoc tweaks to accommodate changes. This lack of adaptivity challenge the efficiency in cost model maintenance.

### C. Graph Neural Networks

When the dataflow graph is too large to hold on the functional unit array, compilers first partition the full graph into subgraphs and then perform placement and routing for each individual subgraph.

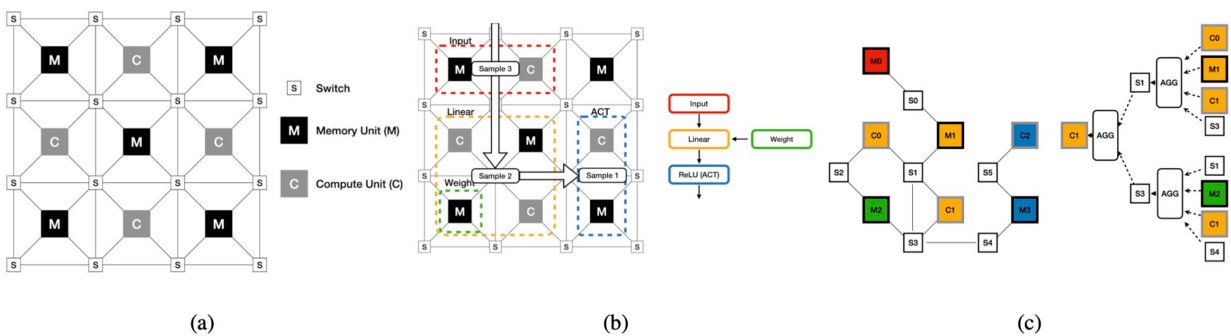

*Fig. 1:* (a) Dataflow architecture: hardware using dataflow architectures consists of a large array of functional units as the data path. Each type of these units can deliver one functionality such as switch, arithmetic compute and memory. Across these units, data are routed using on-chip interconnect. (b) Placement and routing (PnR): To train a DNN on the dataflow architecture, a compiler spatially places a set of arithmetic operations in the DNN simultaneously to the array of units and route the data with on-chip interconnect. Data samples for DNN training are processed in a pipeline fashion where each pipeline stage consists of one or more operations. (c) Cost model based on graph neural networks (GNNs): The PnR decisions induce a graph representation with the units as nodes and interconnect as the edges. To construct a cost model on this representation, we use a regressor based on GNNs to aggregate information through graph neighborhoods and predict the fitness of PnR decisions.

rules typically model the local behavior of units for a standalone operation without considering the interactions between them; these heuristics simplify the modeling. However, they can significantly hold back industrial compiler engineering in three important aspects. Firstly, writing the heuristic-rules with acceptable precision to cover the diverse set of operations involved in training DNNs is an incredibly time- consuming effort. Secondly, the imprecision of heuristics can preclude PnR decisions which are empirically performant but two discouraged under simplified heuristics. For example, operations may share the shortest path for communication on the fabric, and each could require the full bandwidth. Even though in the dataflow graph context they could time-share the routes at runtime, some conservative heuristics could seek to prevent route congestion and encourage a longer route for one of the operations. Lastly industrial compilers are evolving daily in engineering practice. When substantial changes go into the compiler stack such as upgrades to the low-level implementation of an operation, the heuristic rules require ad-hoc tweaks to accommodate changes. This lack of adaptivity challenge the efficiency in cost model maintenance.

### C. Graph Neural Networks

To break the limitation of heuristic-based cost models, we propose a data-driven cost model using machine learning to choose a technique. Towards this end, appropriate machine learning model architecture to encode PnR decisions. We note that PnR decisions directly cast to a graphical representation, with functional units and interconnect as the nodes and edges respectively. Therefore, it is natural to use graph neural networks (GNNs) which are the de facto machine learning tool to model graph-structured data [14]. In Algorithm 1, we present the process to generate the graph-level global representation for input graphs. More formally, let G = (V, E) be a graph where V is the set of nodes and E is V and any edge e the set of edges. For any node

---

**Algorithm 1** Generating graph-level representations

1: Embeddings $\mathcal{X}_V = \{\mathbf{x}_v, \forall v \in V\}$, $\mathbf{x}_E = \{\mathbf{x}_e, \forall e \in E\}$
2: Weight matrices $\mathcal{W}_V = \{\mathbf{W}_V^k, \forall k \in \{1, 2, ..., K\}\}$
3: Weight matrices $\mathcal{W}_E = \{\mathbf{W}_E^k, \forall k \in \{1, 2, ..., K\}\}$
4: **procedure** GRAPHREPRESENTATION($G, \mathcal{X}_V, \mathcal{X}_E, \mathcal{W}_V, \mathcal{W}_E$)
5:     $\mathbf{h}_v^0 \leftarrow \mathbf{x}_v, \forall v \in V$, $\mathbf{h}_e^0 \leftarrow \mathbf{x}_e, \forall e \in E$
6:     **for** $k = 1, 2, ..., K$ **do**
7:         **for** $v \in V$ **do**     ▷ Get node representations
8:             $\mathbf{h}_{\mathcal{N}_{V \to E}(v)}^k \leftarrow \text{AGGR}(\{\mathbf{h}_d^{k-1}, \forall d \in \mathcal{N}_{V \to E}(v)\})$
9:             $\mathbf{h}_{\mathcal{N}_{V \to V}(v)}^k \leftarrow \text{AGGR}(\{\mathbf{h}_u^{k-1}, \forall u \in \mathcal{N}_{V \to V}(v)\})$
10:            $s_v^k \leftarrow \text{MAX}(\mathcal{W}_E * \text{CAT}(\mathbf{h}_{\mathcal{N}_{V \to E}(v)}^k, \mathbf{h}_{\mathcal{N}_{V \to V}}))$
11:            $\mathbf{h}_v^k \leftarrow \mathbf{W}_V^k \cdot \text{CAT}(\mathbf{h}_v^{k-1}, s_v^k))$
12:         **end for**
13:     **end for**
14:     **return** $\mathbf{h}_G \leftarrow \text{AVG}(\{\mathbf{h}_v^K, \forall v \in V\})$
15: **end procedure**

---

### III. DATA-DRIVEN COST MODELING FOR PNR

As discussed in Section II-B, the limitations of conventional PnR cost modeling can hold back the quality of compiler engineering. These limitations intrinsically originate from hand-crafted heuristics in the cost models. To alleviate these issues, we propose a data-driven cost model without heuristics for PnR in compilers for dataflow architectures. Specifically in Section III-A we extract graph representations of the PnR decisions using graph neural networks, the natural machine learning tool for graph-structure data as discussed in Section II-C. Built on the graph representation of PnR decisions, we introduce a regression model to predict and rank the hardware throughput of unseen PnR decisions for unseen dataflow graphs in Section III-B. This GNN-based throughput regressor is trained purely on the empirical throughput of dataflow graphs compiled with randomly generated PnR decisions; thus, it could alleviate limitations from heuristics.

### A. Graph Representations for PnR Decisions

The PnR decisions for dataflow architectures naturally incudes a graph **G** = (**V**,**E**). In such graphs, **V** is the set of nodes containing actively used functional units and **E** is the set of edges representing the used fabric routes. To leverage the GNNs to extract the graph-level representation for PnR decisions, we first construct the node and edge embedding vectors encoding hardware characteristics of the units and fabrics. For a function unit v ε **V**, we define the node embedding vectors as $X_v = [X_{v,F}, X_{O(v)}, X_{s(v)}]$. In this definition first portion of the vector is a one-hot vector encoding the type of the functional unit type, the mapping is the next piece of the representation vector is the learnable embedding vector that returns the type of index of the operation mapped to a unit. The last component in the node embedding vector is another learnable vector $X_{s(v)}$ where $S(.): V \to N$ return the pipeline stage index of a function units. Intuitively $X_{s(v)}$ can partially suggest the order of operations and hint on how many samples can be processed simultaneously by the pipeline, which is relevant information to objectives such as throughput prediction. Finally, to encode characteristics of the fabrics, we define the edge embedding as a fixed vector $X_e$; in this vector, we store features such as the fabric route length associated with e ε E which can suggest how fast data communicates along e. Next, we use the K layer information fusion network in the Algorithm [1] to aggregate the information from $X_v$ and $X_e$ through graph neighborhoods. This aggregation process generates the final node and edge graph $h_v^K$ and $s_v^K$. Taking these vectors and POOL function finally produces a graph-level representation hG. Using this graph-level representation, we can learn a regressor to predict and rank the hardware throughput attain by the DNNs compiled with different PnR decisions.

### B. Throughput Regressor on Graph Representations

To enable our cost model to predict throughput resulting from P R decisions, we consider learning a regression model over the graph-level representation hg. The regression net- work is a simple 3-layer multi-layer perceptron using ReLU as the nonlinear activation function. In order to prepare the data for learning the regressor, we collect a large set of randomly generated PnR decisions G = { $G_g$, ∀ g ∃ {1,2, … N}} from different dataflow graphs. We then measure the training throughput of the artifacts compiled with these randomly generated PnR decisions. We collect the measurements as Y = { Gg, ∀g ∃ { 1 , 2, …, N ) } where ∀ g ∃ R is the consequent throughput of PR decision $G_g$. Using the throughput measurements, we learn the regressor together with the fusion network and embeddings in an end-to-end fashion. The Adam optimizer is used to perform parameter updates [5].

This throughput regressor can be combined with any cost- model-based algorithms to perform placement and routing. Our throughput regressor is lightweight and could be used as a drop-in replacement in production-level compilers to accelerate real DNN training workloads.

### IV. EXPERIMENTS

To validate that our data-driven cost modeling can resolve the pain points in heuristic-based cost modeling, we empirically evaluate the throughput regressor in Section III as a cost model for placement and routing in compiling DNNs to the dataflow architecture. In this section, we demonstrate that when compared to heuristic-based cost modeling, our data-driven approach can eliminate the time cost of manually crafting analytical models using the heuristic, achieve better precision and demonstrate stronger adaptivity to compiler stack changes to maintain throughput advantages in compiling large DNNs. Specifically, we briefly present the experiment setup in Section IV-A and discuss the experiment results in detail in Section IV-B.

### A. Experiment setup

a) Dataset generation: In order to generate the dataset to learn our data-driven cost model, we collect PnR decisions on compiling DNN building blocks, including Generalized Matrix Multiplication (GEMM), Multilayer Perceptrons (MLP), Multiheaded Attention (MHA) and Feed Forward Network (FFN) with various width and depth; these are building blocks that cover a comprehensive list of operations in modern DNNs.

To generate a diverse dataset, we randomized the search parameters of a simulated annealing placer. Regarding the throughput measurement, we observe that the absolute throughput on different dataflow graphs can vary significantly in magnitudes which leads to unstable learning.2 To learn stably, we normalize the absolute throughput into the range [0, 1] with the theoretical performance upper bound. To derive this limit, we simply consider the required amount of compute and the FLOPs for the compute units in each pipeline stage. We then use the limit on the theoretically slowest stage to normalize the absolute throughput measurement; this derivation does not involve any complex heuristics which are time-consuming to craft. In other words, we collect data with the goal of letting our cost model learn to encourage PnR decisions that reach simple theoretical limits. In this dataset generation process, we collect 5878 pairs of PnR decisions and normalized throughputs in total and train a single throughput regressor model on the collected dataset.

b) Heuristic baselines: We use the heuristic-based cost model derived from the expert knowledge on the dataflow architecture as the baseline. In this cost model, each individual operator types have its own rule-based system to capture how fast this operator generate outputs in isolation. A graph-level heuristic predicts normalized throughput and estimates routing congestion from these speed metrics. It takes a large engineering team to develop and refine these heuristics. To measure the quality of different cost models, we use the relative error (RE) to quantify how well a cost model predict normalized throughputs. We additionally use the Spearman rank correlation to evaluate on the capability to rank PnR decisions regarding how close they are to the theoretical limits. To report statistically meaningful metrics, we use 5-fold cross validation across our experiments.

### B. Results

Using the above experiment setup, we compare our data-driven cost model to the heuristic baseline on three aspects which are critical to the compiler engineering practice.

a) Time to construct the cost model: While the baseline requires significant engineering efforts over a long-time span, our data-driven cost model takes approximately one day to construct without intensive human intervention. Specifically, the dataset generation process only takes less than one day in an industrial level CPU compute farm. Upon collecting the dataset, learning the throughput regressor only requires approximately 2 machine hours with one GeForce RTX 2080Ti GPU. This validates that our data-driven cost model can eliminate the significant engineering time cost in constructing heuristic-based cost models.

The absolute throughput is measured by counting machine cycles.

*Fig. 2*: GNN shows to be significant more accurate than heuristic baselines in terms of Rank Correlation and RE.

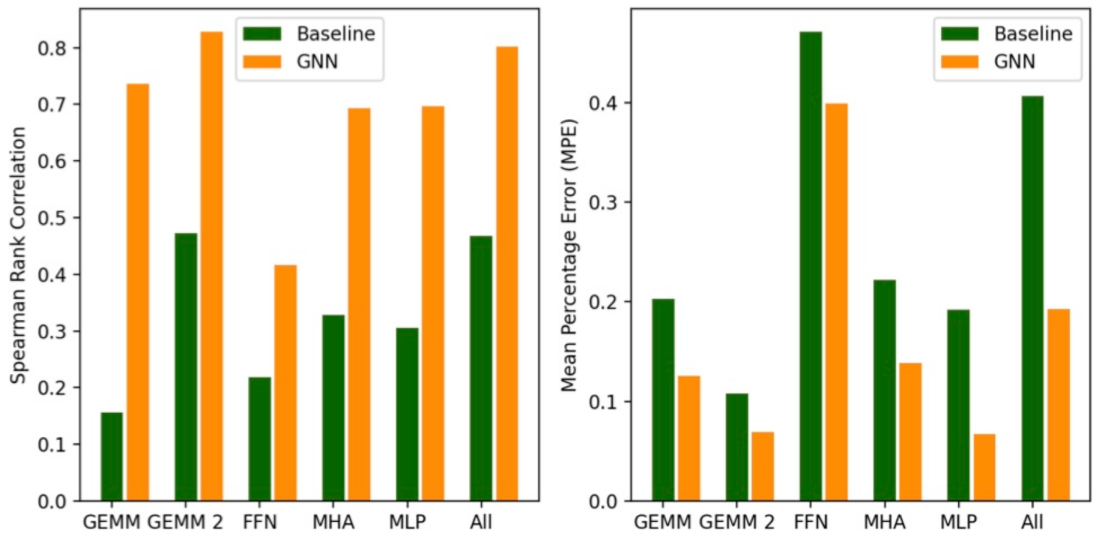

TABLE I: Our GNN beats the baseline significantly in both Relative Error and Spearman Rank correlation in predicting the throughput on a PnR decision

|          | Test RE | Test Rank |
|----------|---------|-----------|
| **Baseline** | 0.406   | 0.468     |
| **GNN**      | 0.193   | 0.808     |

b) Modeling precision: Regarding the precision of cost models, we demonstrate that our data-driven cost model can more precisely predict normalized throughput and rank PnR decisions in terms of normalized throughput. We then empirically show that these advantages can translate to hardware throughput advantages over heuristic baselines in compiling large DNN models. Concretely when comparing predicted normalized throughput to the ground-truth, lesser relative error indicates better prediction precision and higher Spearman rank correlation suggests stronger ranking capability. As shown in Figure 2, across all individual groups of DNN building blocks, our data-driven cost model demonstrates up to 58% higher Spearman rank correlation than the baseline. Similarly, when combining the datasets, we observe 21% improvement in relative error and 34% improment in Spearman rank correlation. Our ultimate goal is to swap the data-driven cost model into the placer and compile high-throughput unseen DNNs. To demonstrate this, we compile several MLP and MHA physical graphs using both cost models in the annealing-based algorithm for graph compilation on the dataflow architecture. We observe that compilations generated with the learned cost model resulted in a 9.1% and 8.6% decrease in latency when compared to compilations generated with a heuristic cost model. For larger and more practical graphs, we compile BERT-large [4] and GPT2XL [12] using both cost models in the same compiler stack. We observe that the BERT-large and GPT2XL compiled with our data-driven cost model can demonstrate 5.7% and 1.3% higher throughput respectively.

This demonstrates the practical use of using the data-driven cost model to generate compilations of logical graphs to increase throughput over several significant model architectures.

c) Adaptivity to compiler changes: Industrial compiler stacks can evolve continuously in timeframes spanning several years. Thus, ease of maintenance is also critical to designing a cost model. Even with substantial compiler stack changes/upgrades, we show that we can easily adapt our data-driven cost model and maintain the throughput advantages over heuristic baseline. In more detail, we recollect the dataset of PnR decisions and normalized throughputs and retrain the throughput regressor at the beginning and end of a time span of 3 weeks, a timespan where 100's of pull requests have affected the software stack. We can observe in Table II that at the two version of the compiler stack, our data-driven cost model can consistently deliver > 5% and 1% throughput advantages over baselines on BERT-large and GPTXL respectively. We have demonstrated that on large and practical logical graphs such as BERT and GPT, our model can generate more accurate predictions of throughput and generate PnR decisions with higher throughput than a heuristic-based cost model despite the large changes in the software behind the Dataflow architecture.

This shows the adaptivity of our model to maintain value over time without the need for large engineering effort to tune the algorithm.

*TABLE II*: Consistent throughput ( TP) and Relative Error improvements on large models by GNN cost model at different timepoints

|       | BERT  |         | GPT   |         |
|-------|-------|---------|-------|---------|
|       | Past  | Present | Past  | Present |
| **RE**    | 0.353 | 0.324   | 0.478 | 0.422   |
| **ΔTP**   | 5.6%  | 5.7%    | 1.1%  | 1.2%    |

## C. Ablation Study

In our data-driven cost model, the node and edge embedding vectors provide foundational information for the GNN to learn a graph-level representation. We perform an ablation study to study the relative importance of node and edge embedding vectors. As evident by Table III, removing the node and edge embeddings causes significant decreases in both accuracy and Spearman rank coefficient over several datasets. This indicates that these features are critical to the success of the data-driven cost model.

*TABLE III*: The decrease in RE and Rank Coefficient demonstrate how the Edge and Node embeddings hold valuable information for the GNN.

|            | RE    |       |       | Rank  |       |       |
|------------|-------|-------|-------|-------|-------|-------|
|            | MLP   | FFN   | MHA   | MLP   | FFN   | MHA   |
| **GNN**        | 0.148 | 0.404 | 0.139 | 0.778 | 0.563 | 0.794 |
| -edge emb. | 0.343 | 0.576 | 0.297 | 0.291 | 0.116 | 0.337 |
| -node emb. | 0.205 | 0.413 | 0.249 | 0.428 | 0.293 | 0.477 |

## V. RELATED WORK

To our knowledge, this is the first study of a data-driven GNN cost model guiding a placement algorithm for a reconfigurable dataflow architecture. Leveraging ML techniques, like GNN, in the cost model design for placement and routing is an active research area under physical VLSI chip design domains [6], [7], [9] but none have extended to the compiler design on reconfigurable dataflow architecture. They also rely on heuristics in the cost model design which causes the aforementioned accuracy problem. In a representative work [6], the authors proposed a similar idea of using embeddings to encode the circuit graph at the transistor level, to predict performance for chip placement. However, there are still three major differences compared to our work. Their graphs were much smaller than ours. The performance labels are obtained through simulations rather than real measurements as in our work. Only the placement is modeled in their work, while we additionally model the routing. In [9], the idea of using a GNN to predict the return in the value network of a RL placer is similar to a cost model. However, instead of training on real measurements, they use a performance estimation proxy for the final reward to reduce the overhead in environment interaction. Another popular PnR algorithm is to formulate the hardware constraints into an Integer Linear Program (ILP) [10]. However, the cost model has to be simple and linear to be used in ILP but PnR on dataflow architecture has many complex hardware constraints.

## VI. CONCLUSION

In this paper, we challenge the conventional wisdom of to guide PnR compilation on using heuristic cost model reconfigurable dataflow architecture. We show that by taking easily accessible hardware features, our data-driven GNN-based cost model can avoid the complex heuristic design phase of traditional cost model and save months of engineering time. We show that our cost model can give better throughput predictions on different DNN building blocks. In addition, our cost model pretrained on these building block can be directly used to compile larger, unseen models such as BERT/GPT and attains up to 5% throughput improvement. We also demonstrate that our cost model is adaptive to changes in the compiler by consistently beating the heuristic baseline in terms of throughput.